# A Deep Look Into - Automated Lung X-Ray Abnormality Detection System

Nagullas KS[1], Vivekanand.V[1], Narayana Darapaneni[2] and Anwesh R P[3,*]

[1] PES University, Bangalore, Karnataka, 560050, India
[2] Northwestern University, Evanston, IL 60208, United States
[3] Great Learning, Hyderabad, Telangana, 500089, India

## Abstract

INTRODUCTION: "Automated Lung X-Ray Abnormality Detection System" is the application which distinguish the normal x-ray images from infected x-ray images and highlight area considered for prediction, with the recent pandemic a need to have a non-conventional method and faster detecting diseases, for which X ray serves the purpose.
OBJECTIVES: As of current situation any viral disease that is infectious is potential pandemic, so there is need for cheap and early detection system.
METHODS: This research will help to eases the work of expert to do further analysis. Accuracy of three different pre-existing models such as DenseNet, MobileNet and VGG16 were high but models over-fitted primarily due to black and white images.
RESULTS: This led to building up new method such as as V-BreathNet which gave more than 96% percent accuracy.
CONCLUSION: Thus, it can be stated that not all state-of art CNN models can be used on B/W images. As a conclusion not all state-of art CNN models can be used on B/W images.

## 1. Introduction

Health is vital to human life, and to ensure healthy life a sound health care system is necessary. The technology progress en- sured advancement in health care system. Timely delivery and detailed analysis of report are crucial in terms of performances of health care system, these attributes can be improved by utilizing the latest development of Artificial Intelligence (AI) thereby leading to sound inspection to diagnose symptoms and identifying diseases.There is always a risk factor of wrongly diagnose of symptoms but on contrary we are not into completely replacing doctor with A.I, but to help them to treat patient better and in short time. A dire lesson learnt from COVID period was that hospital staff were severely understaffed to handle a pandemic and time was of essence. Higher workload influences human error due to various factors influencing, as such in a pandemic a system that could help reduce the staff to diagnose the disease at cost-effective rate is needed. For COVID and Pneumonia medical photographs such as CAT scan, X-Ray are significantly [1] useful to doctors or AI for detection of so said disease.

Around the world there have been more deaths due to Pneumonia than AIDS, in latest pandemic we can see SARS category related viral disease COVID-19 had caused higher death rates. With possibility of mutation of the virus current test of PCR is currently in use and this is time consuming and costly for many of economical backward countries. Now we have portable x-ray system that can extract x-rays remotely this can be a cheaper alternative, this will also be better option for patients who might be disabled or can't make it to testing centre [2]. By using X-Ray of chest area pneumonia or covid is able to be detected with high accuracy would relief stress on the health care system to certain extent as an early detection. Our goal was to develop a highly efficient Automatic X-Ray classification system from CNN based architecture that can distinguish between COVID-19, viral pneumonia, and normal lungs, and is easily retrainable for different virus patterns, while also being deployable in low-power legacy medical devices for accessibility in lower-income countries taking less space than current state-of-art models

## 2. Related Works

Pneumonia is a viral disease that can become deadly if not detected at right time and treated appropriately. There are few ways to diagnose pneumonia for example x-ray, CT scans, ultrasound scans of the chest and few other type of scans of the chest can all be utilized. X-rays of chest is preferred [3] more than CT scan as it will be much cheaper, faster and is widely available diagnostic imaging technology. There is lack of competent healthcare staff and radiologists across the world, whose ability to predict such illnesses is crucial. Features of multiple aliments overall lap onto each other and human eye struggles to identify small patterns [4] that could differentiate different type of diseases.

Deep learning algorithm have increasingly become popular in field of medicine as it is able to identify minute of pattern that average human eye struggles to do so. In one of studies, they found that the application of deep learning and computer vision algorithms in biomedical picture diagnosis was extremely beneficial in terms of giving a speedy and accurate diagnosis of disease, which was comparable to the accuracy of a radiologist. Deep learning-based algorithms are currently unable to replace competent clinicians in medical diagnosis, and they are meant to be used to supplement clinical decision-making rather than to completely replace it.

Chest radiographs play a role in diagnosing COVID-19 pneumonia, but they have limitations. Normal chest radio- graphs show a clear central mediastinum and heart, air-filled lungs appearing black, present lung markings representing blood vessels, and a curvilinear diaphragm with sharp margins [5]. However, in COVID-19 pneumonia, certain features may be observed. Ground glass opacity, an initial abnormality, appears as increased whiteness on chest radiographs, giving a ground glass appearance [2]. Radiologists may confirm this finding. Horizontal linear opacities can also be seen alongside ground glass changes. Ground glass opacities are often bilateral but can be unilateral, primarily affecting the peripheral lung regions, especially in the mid and lower zones. As the disease progresses, ground-glass opacities become denser and progress to consolidation, leading to a com- plete loss of lung markings. Consolidation areas are likely to have progressed from sites of ground glass opacities. However, chest radiographs have limitations. Anteroposterior (AP) images from portable machines produce lower-quality images compared to posteroanterior (PA) chest radiographs from dedicated radiography facilities, making interpretation more challenging. AP chest radiographs may also magnify the appearance of the heart. CT would be much suited in the diagnosis of COVID-19 infection for detection on early signs
of disease and may be more sensitive than chest X-ray [1].

Combination of RCNNs and long short-term memory (LSTM) networks was used [6]. similarly a model was called as VDSNet [7]. The CNNs are used to extract features from the X-ray images then connected a pre-trained VGG16. The VGG16 majorly focused on to capture increasingly complex patterns in the images. The output from the VGG16 network is then fed into a set of LSTM networks, which are used to analyze the sequences of features extracted from the CNNs over time. There was another model used for object detection using solely virtual world data. They simulated environment to generate a large-scale dataset of annotated images, which they then use to train a CNN with a modified YOLOv2 architecture [8].

CNN based network to detect pneumonia disease only with VGG19 model as a base model with transfer learning with data augmentation making more than 12000 images total was trained. They proceeded with advanced technique called "Deep Convolutional Generative Adversarial Network" [9]. The im- ages in the dataset were annotated by utilizing both metadata and the actual content of the images. This was accomplished through the implementation of a technique called Content- based Image Retrieval. Model contains 54 million params and with 300 epohs 97% accuracy was observed.

The LSTM networks in the hybrid model are trained using a technique called transfer learning, which involves initializing the network weights using a pre-trained model and fine- tuning them on the specific task at hand, evaluated their hybrid deep learning model for pneumonia, tuberculosis, and lung cancer, as well as healthy controls. They found that the model achieved high accuracy in detecting these diseases, with a validation accuracy of 73 for complete dataset training. Overall, the hybrid deep learning model described in the paper is a sophisticated and effective approach to analyzing X-ray images of the lungs for the purpose of detecting various lung diseases.

A 3D CNN variant called the ResNet-50 architecture, which has been pre-trained on large-scale image recognition datasets to improve the efficiency of training the classifier was utilized to extract features from the lung images. The 3D CNN extracts features from the CT scans by analyzing the spatial informa- tion across the three dimensions (x, y, z) of the CT scans [10]. The extracted features are then used to train a classifier to distinguish between COVID-19 positive and negative cases, provided 84% accuracy. In other study propose a new method that uses CNN to learn the camera's perspective from 2D images of the human body [11]. The CNN is trained to predict the camera's angle of view which minimizes errors in 3D body posture estimation

The CoviSegNet [12] is trained on a dataset of lung CT images using a combination of supervised and unsupervised learning techniques. The supervised learning is used to train the model on a labelled dataset of COVID-19 infected lung images, while the unsupervised learning is used to fine-tune the model on an unlabelled dataset of lung images. Inception model [13] that will be more of intermediate step -in between regular convolution and the depth wise separable convolution operation. Based on single-shot detector RetinaNet with Se- ResNext101 encoders, pre-trained on ImageNet dataset. With extra improvisation involved to improve accuracy was effective boosting accuracy by greater margin.

## 3. Materials And Methods

### 3.1 Data Pre-processing

Table 1 SIRM Covid-19 Dataset Details

| Property | Value |
| --- | --- |
| Total Instances | 15 thousand images |
| Size | 1.12 GB |
| Category | 3 types |
| COVID images | 3616 instances |
| Normal images | 10192 instances |
| Pneumonia images | 1345 instances |

Data used for our research is taken from "Italian Society of Medical and Interventional Radiology (SIRM) COVID-19" dataset contains black and white lung x-ray images and it has divided into three category such as normal, covid and pneumonia. Data consists of COVID images of 3616 instances Normal images of 10192 instances Pneumonia images of 1345 instances as seen in table I. Data is unbalanced in terms of number of instances here it was addressed by having data augmentation to bring up data instances of existing images to desired numbers. Prior to which high and very low contrast images were removed in-order to retain images with most visible pattern as well as images towards normal x-ray. Blurry images were also handled using variance of Laplacian meaning that roughness or high sharpness of image is used to look at changes in image again a threshold and images with lower variance of sharpness indicated blurry images, hence they were removed. After all said pre-processing was done we had 3200 images of each alignment type image which was split to train and test data for model instances as shown in table 2.

Table 2 - Training And Testing Dataset Details

| Training Set | Testing Set |
| --- | --- |
| 2560 X-Rays COVID-19 | 640 X-Rays COVID-19 |
| 2560 X-Rays COVID-19 | 640 X-Rays COVID-19 |
| 2560 X-Rays COVID-19 | 640 X-Rays COVID-19 |
| Total: 7680 images | Total: 1920 images |

### 3.2. Experimentation-Transfer learnt CNN models

Convolutional neural networks (CNNs) generally have the ability to directly process raw images without any prior preprocessing, when a feed-forward neural network consists of higher number of layers. The convolutional layer is a crucial component that empowers CNNs. Imagine a convolutional layer as a set of small square templates, known as convolutional kernels, that move across an image in search of patterns.. Many convolutional layers are piled on top of each other in convolutional neural networks, each one capable of detecting more complex shapes. If we combine this power of CNN with other method called transfer which leverages knowledge gained from pre-trained models saving learning time for smaller dataset. This approach saves time and computational resources required to train a model from scratch. Moreover, transfer learning is especially useful in situations where the size of the training dataset is limited, as it enables the model to learn from the patterns and features present in larger datasets. The pre-trained models can be fine-tuned by either retraining some or all of the layers or by freezing certain layers and training only the remaining layers. Experiment was conducted taking advantage of transfer learning of state-of-art models like VGG16, DenseNet and MoblieNet V2. Weights from ImageNet data set was utilized as initial weights, this would serve an advantage for training model. Experiment was carried on said models in google colab environment for couple of rounds in order to get more consistent readings for the model performance to learning the features of the x-rays. The dataset was pre-processed to remove all the blurry images, contrast was set properly, data augmentation was also conducted to boost the learning capabilities of the model. Data was balanced with respect to available categories to maintain equality in data to learn and not be biased in the model. Post that dataset was split to train and test with 80-20 ratio respectively. This allowed sufficient data for to be tested and same structure of data was utilized across the models for consistent reading. Model was trained with learning rate of 0.01 as this was found to be ideal learning rate for the model after training model for different learning rate. Epoch was set to 5 due to constraint of the colab getting disconnection or GPU usage limit reach potentially making the model to stopping in between. The VGG16 was unable to perform in time frame of the GPU usage and had to be run from the normal run-time provided by google colab. The findings from the experiments indicates on the model performing well initially and could be seen to have around 92% validation accuracy but with trade off of model over-fitting later epohs.

Table 3 - Represents Metrics Of Experiment And Observation

| Models | DenseNet | MobileNet-V2 | VGG16 |
| --- | --- | --- | --- |
| Total parameters | 8,217,667 | 3,733,059 | 15,305,027 |
| Trainable parameters | 1,180,163 | 1,475,075 | 590,339 |

| Maximum epochs | 5 | 5 | 5 |
|---|---|---|---|
| Max Val Accuracy | 92.24 % | 92.60 % | 92.75 % |
| Maximum Loss | 0.1767 | 0.1730 | 0.1780 |
| Training Time | 45 mins | 49 mins | 8.5 hours |

DenseNet and MobileNet models exhibit faster training times compared to VGG16 as VGG has much depth making it computationally expensive III. Even in terms of parameters VGG16 leads, however the validation accuracy of all the models for 5 epochs is similar. Imparting that transfer learned models dataset may not be ideal for black and white images like x-rays. Model size also plays a role, as excessive parameters can lead to over-fitting instead of effective learning, poten- tially due to the significant size difference between ImageNet and the dataset used in this experiment. The maximum loss observed for all the models is around 0.17, indicating that the models have reached or are close to the minima. However, the models tend to exhibit over-fitting beyond the initial epochs, as evidenced by the difference between training and validation accuracy. The experiment was limited to 5 epochs due to the constraints of using free GPU resources on Google Colab. In the medical field, data scarcity is a common challenge, as obtaining sufficient and well-labeled data is expensive and requires specialized equipment and trained personnel.

### 3.3 Implementation

As over-fitting became major problem from transferred learnt state-of -art models we have to proceed to build a model from freshly that performs on par or exceeds the accuracy seen by the previous experiment results. Convolutional neural network technology was continued for new model as well and was built on Google Colab with keras package. Model structure was following same pattern as in convolutional layer followed by a batch normalization, max-pooling layer and finally dropout layer, the batch normalization was carried out to improve consistency of the data between convolutional layer and overall improve the learning process. Max pooling was done to reduce the dimensionality and to retain main features in the model, this was done along with padding to retain most important features properly so as to improve model learning. Dropout is technique used for helping model to reduce over- fitting by dropping a unit in neural network with predefined probability, this forces other units to learn making model not to over-fit to a certain degree.

The last convolutional layer is processed by a fully connected layer with 512 neurons followed by last three dimensional neurons representing individual x-ray stage. Activation function utilized in the model is ReLU while softmax was used to tome of final dense layer as for multi classification. Filters in convolutional layers were varied, spanning from 128 to 32 and back to 128 becoming as closer match for letter 'V' hence the name V-BreathNet. In order to reduce overall parameters number of filters had to be reduced before flatting it and 64 filter was found to be most suitable to retain major of features and have decent parameters.

The data was split into 80:20 ratio making 7608 images for training that includes all three categories of x-ray equally distributed, 1902 images were utilized for validation of the model. Categorical cross-entropy was used for loss calculation as its suited for the multi-class classification, optimizer that helps in update of models weights was set to rmsprop, this allowed optimizer to adapt to different features in the data thus helping to accelerate convergence and improve training efficiency.

Another important factor was considering the initial learning rate which was 0.001, we also made sure to reduce the learning rate by 0.5 after every 5 epochs to prevent the accuracy metric from plateauing over a particular range. This helped in achieving required accuracy more efficiently, minimum threshold was set to 0.000001.Overall the model took fairly decent time to complete training. Model consisted of 792,291 parameters which is less than one third of the MobileNetV2 which had least amount of parameters among experimented models.

Pattern seen by the CNN model does it match to features what radiologist look into to investigate and understand model in better terms we utilized saliency map to highlight areas of the input image that model was utilizing most in order to predict. we applied GradCAM (Gradient-weighted Class Activation Mapping) we were able to generate heat maps that emphasized the areas of the input image that were most significant for a particular class, post over lay of the areas identified by saliency map on original image we were able co-relate on actual effected area. This visualization technique allowed us to interpret and validate the model's decision-making process, as well as identify the important regions that contributed to its predictions.

### 4. Results

As our first approach was to proceed with state-of -art mod- els such as DenseNet, VGG16, MobileNetV2 to get transfer learnt some of features that was trained with ImageNet dataset and proceed to train SIRM dataset. But with this experiment along with limited resources we observed that model was overfitting as the accuracy during training exceeded validation accuracy for epohs less than 5. Also, models are leaning towards high parameters ranging more than 5 million params. Due to these high parameter numbers model can over-fit as dataset pales in comparison with sheer number of images in ImageNet data set.

Now we knew custom model needs to be in place which can yield better accuracy while maintaining parameters to lower end and experimentation to determine number of layers, filters etc started. After going through couple of base structure and filter we finally settled with V-BreathNet architecture and proceeded to training. Training of model was complete and we observed better accuracy without model over-fitting. Highest validation accuracy seen from model during training was 96.84%. Model is able to predict whether given x-ray has pneumonia, covid-19 or is normal x-ray, but as in any

A Deep Look Into - Automated Lung X-Ray Abnormality Detection System

other model case there were some x-rays wrongly classified justifying the accuracy to be as mentioned.

The testing of model was carried out with 90 images comprising of each classification equally and we had observed that normal, viral pneumonia x-rays were accurately identified, but covid x-rays had few wrong classifications 25 out of 30 images were correctly identified initially. Upon further study of the images, we had identified that lung area was lesser compared to correctly identified images, post adjusting image to focus more towards the chest area and not on diaphragm and below area. Post image cropping to focus on chest area more 2 out of 5 wrongly identified was rectified to covid classification.

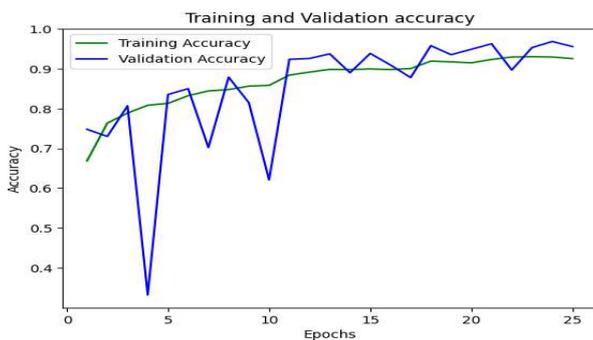

Figure.1. Training and validation accuracy of V-BreathNet.

As we can see from figure 1 first couple of epohs itself the accuracy had a drop and in further epohs as well there were drops in the accuracy as model was learning and had overshot the minima with that epohs learning rate and were corrected in further epohs by loss function even the loss behavior justifies same as seen in figure 2. At 24th epohs the highest peak in accuracy was achieved with validation accuracy being 96.84% and validation loss was at 0.0964.

We utilized the saliency map produced heat maps in order to research more on why model wrongly classified and found out that the patterns in falsely identified as normal x-ray are more towards the edges of the lung area, another majorly area utilized area for identification was diaphragm area which is indicating that the patients might have been in initial stages.

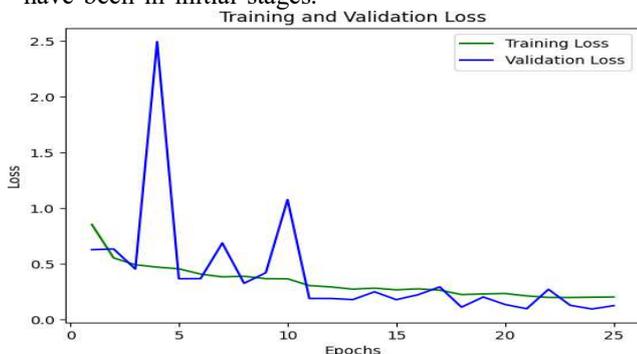

Figure. 2. Training and validation loss of V-BreathNet.

There is lack of different stages x-ray images in enough numbers for model in understand better. Once infection is spread and is visible enough in x-ray it is effectively identified also there is significant impact on diaphragm in long exposure to the covid, mechanical breathing support on diaphragm [14] which can contributor pattern in identifying covid.

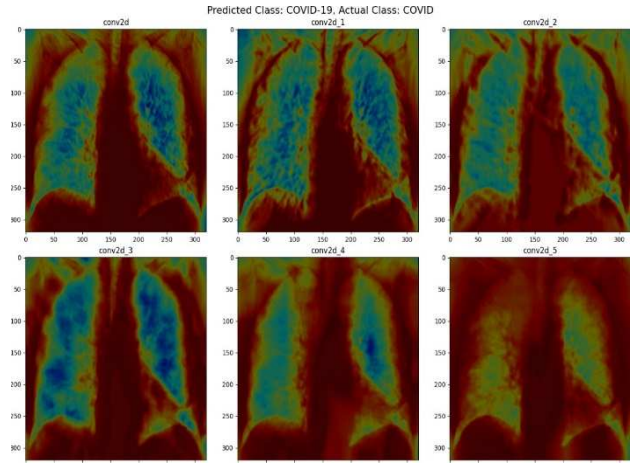

Figure.3. Saliency map of Covid patient x-ray predicted by V-BreathNet.

V-BreathNet first convolutional layer is starting to identify patterns are outer edge of lungs indicating edge nodes of lung. One of other pattern seen by model is the heart region and diaphragm showcases that these regions are impacting prediction. As in deeper layer the opacity of lung is taken into pattern identified by the model hence in over-layed heat map, we can see the patterns reaching from outer edge of the lung to inner section of the lungs. Normal x-rays are more coverage in cardiac, diaphragm, shoulder region and major contributor would be edges of rib cage. Pneumonia and Covid diseases have impacting area on the outer edge as see in figure 3 as well as the left lung seems to be affected more for majority of x-rays considered for testing. Performance metrics of the model is as seen in table 4 .

Table 4 - V-BREATHNET PERFORMANCE METRICS

| Class | Precision | Recall | F1-Score |
|---|---|---|---|
| Normal | 0.94 | 1.00 | 0.97 |
| Covid-19 | 1.00 | 0.90 | 0.95 |
| Pneumonia | 0.97 | 1.00 | 0.98 |

The precision values of 0.94, 1.00, and 0.97 for classes normal, covid, and pneumonia respectively indicate that model predicts a sample belonging to a particular class is correct with high accuracy. The recall values of 1.00, 0.90, and 1.00 signify that the model successfully captures a high proportion of the true positive samples for each class, minimizing the number of false negatives. The F1-score, which combines precision and recall into a single metric, shows a high performance for all classes (0.97, 0.95, and
5

0.98), indicating a good balance between precision and recall.

## 5. Discussion And Conclusions

Transfer learning is much preferred than making model to learn things from scratch but it has its own limitation one of which is seen at beginning of research, experimentation of this research. Over-fitting was major issue in terms of model accuracy, due to nature of model used for transfer learning they were not suitable for lesser processing capacity archiving medical applications considered for lower income countries. One of biggest of aspect was large amounts of parameters involved in models even though they were a necessity for accurately predict multi-classification on a large dataset such as ImageNet dataset, but that itself backfired for a small dataset such as SIRMS x-ray dataset.

To overcome these limitations, we developed the V-BreathNet architecture with contained much lesser parameters than MobileNetV2 which was consisting least among 3 models considered for transfer learning experiment. The outcome from research demonstrated that the V-BreathNet model outper- formed the transfer learned models, achieving a validation accuracy of 96.84%. The precision, recall, and F1-scores for each class showed high performance, indicating accurate predictions for normal, COVID-19, and pneumonia cases. The model exhibited better generalization and reduced over-fitting compared to the pre-trained models. Also satisfying condition to occupy lesser space and run in fairly basic systems to support economical backward country computation abilities.

In order to gain insights into the model's decision-making process we had utilized heat map generated by saliency maps under GradCAM. These heat maps highlighted the regions of the input images that were most significant for classification. The analysis revealed that the model focused on lung edges, ar- eas of opacity seen and cardiac regions which aligns to certain extent with radiologist approach to read x-ray for classification [2]. However, some miss-classifications occurred, particularly in COVID-19 cases where lung area was less visible. Adjusting the image focus on the chest area improved the accuracy of classification. There is still significant need more diverse x- ray images for covid-19 and x-ray that can be indicate early stage of covid infection would provide more insight on which area is primarily hit, also on how lung changes in terms of x-ray image. That would potentially help the model to leap to greater accuracy along with high retain-ability, model would be much more immune to bias due to training data.

Overall, the V-BreathNet model showcases the significance of developing custom architectures tailored to specific datasets and tasks. Its performance metrics, combined with its interpret ability using saliency maps, contribute to the understanding and trustworthiness of the model's predictions. This research makes a helps to better understand use case of AI in contribu- tion to the field of medical image classification, aiding in the early detection and accurate diagnosis of respiratory diseases.